\documentclass[useAMS]{mn2e}
\usepackage{times}
\usepackage{rotating}
\long\def\symbolfootnote[#1]#2{\begingroup%
\def\thefootnote{\fnsymbol{footnote}}\footnote[#1]{#2}\endgroup}
\newcommand{\gae}{\lower 2pt \hbox{$\, \buildrel {\scriptstyle >}\over {\scriptstyle
\sim}\,$}}
\newcommand{\lae}{\lower 2pt \hbox{$\, \buildrel {\scriptstyle <}\over {\scriptstyle
\sim}\,$}}

\begin{document}

\title[Adiabatic Expansion and the central engine in GRBs]
{Adiabatic expansion, early x-ray data and the central engine in GRBs}

\author[Barniol Duran \& Kumar]{R. Barniol Duran$^{1,2}$\thanks
{E-mail: rbarniol@physics.utexas.edu, pk@astro.as.utexas.edu}
and P. Kumar$^{2}$\footnotemark[1] \\
$^{1}$Department of Physics, University of Texas at Austin, Austin, TX 78712, USA\\
$^{2}$Department of Astronomy, University of Texas at Austin, Austin,
TX 78712, USA}

\date{Accepted  2009 February 1;
      Received 2009 January 23;
      in original form 2008 June 6}

\pagerange{\pageref{000}--\pageref{000}} \pubyear{2008}

\maketitle

\begin{abstract}
The {\it Swift} satellite early x-ray data shows a very steep decay
in most of the Gamma-Ray Bursts light curves.
This decay is either produced by the rapidly declining continuation of the 
central engine activity or by some left-over 
radiation starting right after the central engine shuts off.
The latter scenario consists of the emission 
from an ``ember'' that cools via adiabatic expansion
and, if the jet angle is larger than the inverse of the source Lorentz factor, 
the large angle emission.  In this work, we calculate
the temporal and spectral properties of the emission from 
such a cooling ember, providing a new treatment for the micro-physics
of the adiabatic expansion.  We use the adiabatic invariance of $p_{\perp}^2/B$ 
($p_{\perp}$ is the component of the electrons' momentum normal to the 
magnetic field, $B$) to calculate the electrons' Lorentz factor during 
the adiabatic expansion; the electron momentum becomes more and more aligned
with the local magnetic field as the expansion develops.  
We compare the theoretical expectations 
of the adiabatic expansion (and the large angle emission) with the current 
observations of the early x-ray data and find that only $\sim 20\%$ of 
our sample of 107 bursts is potentially consistent with this model.  
This leads us to believe that, for most bursts, the central engine does 
not turn off completely during the steep decay of the x-ray light curve; 
therefore, this phase is produced by the continued rapidly declining 
activity of the central engine.

\end{abstract}


\section{Introduction}

The central engine of the GRBs (Gamma-Ray Bursts - see Piran 2005 
for a review) is hidden to direct observations and its workings
are largely unknown.  
The only information that we currently have about the GRB is
obtained from its electromagnetic radiation.  We have to look for 
signatures in the radiation mechanism to understand 
how Nature produces these outbursts.

{\it Swift} has provided very early x-ray data that 
shows that for most bursts there is a very steep decay
lasting for about 10 minutes 
(Tagliaferri et al. 2005,  Nousek et al. 2006 - see Zhang et al. 
2006 and references therein for possible physical explanations).  
These observations suggest that the rapidly declining x-ray 
light curve (LC) and the burst are produced by the 
same source, because the x-ray LC, when extrapolated
backwards in time, matches the gamma-ray LC (O'Brien et al. 2006).  
Therefore, a natural 
question arises: Is the early x-ray data really just a
rapidly declining 
continuation of the central engine activity, originally 
seen in the gamma-ray band, but now seen at lower energy? 
Or does the central engine switch off abruptly and the early 
x-ray data doesn't reflect the activity of the central engine?

If the central engine completely shuts off when the 
gamma-ray photons' flux falls below the gamma-ray 
detector sensitivity,  
then the emission from a  cooling ``ember'' would be 
responsible for the early x-ray steep decay.  This  
source - which had just produced the gamma-ray emission - 
would be cooling by adiabatic expansion (AE).  
In this paper, we study
the flux properties of a ``hot'' shell that undergoes
AE and cools.  The goal is to determine if the 
observed x-ray steep decay
data is consistent with the AE scenario.  If so, then 
the central engine did shut off abruptly right after the 
gamma-ray emission ceased.  On the other hand, 
if it isn't, then the data is reflecting the rapidly 
declining activity of the central engine.  The reason 
for this is that any other process that does not invoke central 
engine activity to explain the early x-ray data has problems
explaining the smooth temporal connection observed 
in the LC between the prompt emission and the early x-ray data.  

AE has been studied before to predict the long-wavelength 
afterglow from GRBs (M\'esz\'aros \& Rees 1997) 
and also, the optical flashes from internal and reverse shocks 
(M\'esz\'aros \& Rees 1999, Sari \& Piran 1999a).  In this work, 
we describe the evolution of a collisionless plasma due to AE
and show that this is in general different from AE of an 
ideal gas.  

We first present the micro-physics of the AE for 
a collisionless plasma in \S2.  
Then, we use it to calculate the flux properties of 
a source undergoing AE in \S3.  We look at what the 
current observations tell us in \S4, and we put them in the 
context of the central engine in \S5.  We summarize our 
results and give our conclusions in \S6.

\section{Micro-physics of the Adiabatic Expansion}

For an ideal gas, the pressure evolves due to adiabatic 
expansion as $P_{ej}\propto \rho^{a_e} \propto V^{-a_e}$, where 
$a_e = 4/3$ for a relativistic gas and $\rho$ and $V$ are the co-moving 
density and ejecta volume, respectively.  The collisions between electrons 
are extremely rare in GRB relativistic shocks, therefore, one needs 
to be careful in the use of this formula\footnote{We have estimated the 
mean free path for Coulomb scattering between a hot electron and a cold 
electron and find it to be much larger compared to the shell thickness.  The 
electric field associated with a relativistic hot electron is calculated 
using the Li\'{e}nard-Wiechert potential, and we find the cross-section 
for a significant interaction, i.e. leading to a fraction of the energy 
of the hot electron transfered to a cold electron, is close to the 
Thomson cross section ($\sigma \approx \sigma_T/3$).  The mean free path is 
$\lambda=(n\sigma)^{-1}$, where $n$ is the electron density.  
From the total energy $E$, the distance of the shell from the center of explosion $R$,
the source LF $\Gamma$, and the co-moving shell width $\Delta'$, 
we find $n=E/(\Gamma m_p c^2 4 \pi R^2 \Delta')$. 
Using the usual notation $Q_n=Q/10^n$, we obtain:
$\lambda/\Delta' \approx 700 R_{15}^2 \Gamma_2 E_{52}^{-1}$.  For scattering 
between hot electrons, the conclusion is the same. Therefore, Coulomb 
scattering between the electrons is not significant.}.

For a collisionless magnetized plasma, assuming that no collective 
plasma processes randomize the particles' velocity due to scattering, 
particles move along a 
magnetic field line and, by using the concept of 
{\it adiabatic invariant} (Jackson 1998, Rybicki \& Lightman 1979),
we can calculate the particles' energy.  This invariant describes 
that, for slowly varying 
fields,  the magnetic flux through the orbit of the particle is a 
constant, or $p_{\perp}^2/B$ is an adiabatic invariant, where
$p_{\perp}$ is the component of the particle's momentum transverse to $B$, 
the magnetic field.  For highly relativistic particles, 
$p_{\perp} \approx m_e c \gamma_{\perp}$, so that 

\begin{equation}
\gamma_{\perp}^2/B = \textrm{constant}
\label{eq:one}
\end{equation} 
can be used, where $\gamma_{\perp}$ is the Lorentz Factor (LF) of the 
electron in the transverse direction, $m_e$ is the 
electron's mass and $c$ is the speed of light 
(from now on, $c=1$).  This relationship can be used 
because the magnetic field decays on a much larger length-scale
than the electron's gyro-radius (see Appendix A).  It is worth noting that the 
parallel component of the electron's momentum remains unchanged; this will 
be briefly discussed in the last section.

In the next sections, we will make use of (\ref{eq:one}) 
to predict the evolution of the electrons' LF in a hot 
shell that undergoes AE.  We will use it to calculate the 
properties of its synchrotron and synchrotron-self-Compton 
(SSC) radiation.  

\section{Analytical light curves of an adiabatically cooling ember}

Let's assume that the GRB ejecta was heated
by some process (shocks or magnetic dissipation) 
and suddenly the central engine 
switches off completely.  There is no other energy 
injection mechanism at hand, 
so it begins to coast (the LF of the ejecta is constant,
see \S3.6) 
and cools via AE.  We will calculate the flux properties 
of this cooling ember. 

\subsection{Ejecta width and magnetic field}

In the following subsections we will provide the basic ingredients for the 
radiation calculation.  First, we need to determine the 
co-moving thickness of the ejecta, which could be: 

\begin{displaymath}
\Delta' = R/\Gamma \qquad \textrm{or} \qquad \Delta'= \Delta_0.
\end{displaymath}  We will call these cases: thin ejecta shell 
(an ejecta that undergoes significant spreading) and thick 
ejecta shell (an ejecta that experiences no significant spreading),
 respectively.  The observed time is $t \propto R/\Gamma^2$,  
where $R$ is the distance of the source from the center of the 
explosion and $\Gamma$ is the LF of the source with respect to the rest 
frame of the GRB host galaxy. 

The magnetic field in the GRB ejecta could be a combination 
of frozen-in field from the central explosion and field 
generated locally.  We prescribe the decay of the field by using 
the flux-freezing condition (Panaitescu \& Kumar 2004), which gives:

\begin{displaymath}
B_{\perp} \propto (R\Delta')^{-1} \qquad \textrm{and} \qquad B_{\parallel}\propto R^{-2}.
\end{displaymath} We will use the field that decays slower, although this is 
highly uncertain.  This is because the magnetic field generation mechanism for 
GRBs is still not well understood, so the relative strength of 
$B_{\perp}$ and $B_{\parallel}$ is unknown.

\subsection{Electrons' energy distribution}

For an adiabatically expanding source, no more energetic 
electrons are injected in the system when the shock has run its course. 
This means that, after some time, there will be 
few electrons with energies higher than the cooling electron 
LF,  $\gamma_c$.  Therefore, the electron population above $\gamma_c$
will be truncated due to radiation losses and the emission for $\nu_c < \nu$
will rapidly shut off ($\nu_c$ is the synchrotron frequency 
corresponding to electrons with $\gamma_c$).  At this point, 
the electron distribution will follow $\propto \gamma^{-p}$ 
for $\gamma_i < \gamma < \gamma_c$, where $\gamma_i$ is the 
typical LF of the electrons,
since we would be dealing only with adiabatic electrons.
Moreover, since the radiative
cooling quickly becomes less important than the adiabatic cooling
because the magnetic field decays rapidly with the 
expansion of the ejecta, both 
$\gamma_i$ and $\gamma_c$ will evolve in the same way.  

For the case $\gamma_c < \gamma_i$, the radiation losses would 
dominate and, after some time, they would make the whole electron distribution 
collapse to a value close to $\gamma_c$. A narrower range in the 
electron distribution would be responsible for the radiation. 
In this paper we focus on the $\gamma_i < \gamma_c$ case.
       
\subsection{Basics of Synchrotron and SSC}

The electrons' four-momentum is given by 
$P =m_e(\gamma, \gamma \cos \alpha',\gamma \sin \alpha',0)$, which 
can be also written as $P=m_e(\gamma,\gamma_{\parallel},\gamma_{\perp},0)$, 
where $\gamma_{\parallel}$ is the component of the electrons' momentum 
parallel to $B$.  The pitch angle, which is the angle between $B$
and the velocity of the electrons, is $\alpha'$.
In this notation, the electron's LF is then 
$\gamma^2 = \gamma_{\parallel}^2 + \gamma_{\perp}^2$.  
According to the prescription of the adiabatic invariance,
$\gamma_{\perp}$ evolves following (1) and $\gamma_{\parallel}$ 
remains unchanged.  We assume that at the onset of the adiabatic 
expansion $\gamma_{\perp} \sim \gamma_{\parallel}$, then 
quickly when time doubles, 
the radius would have also doubled, making the magnetic field 
decrease by at least that factor and reducing 
$\gamma_{\perp}$ making $\gamma_{\parallel} > \gamma_{\perp}$, 
which gives:

\begin{equation}
\gamma = \gamma_{\parallel} \sqrt{1+ \frac{\gamma_{\perp}^2}{\gamma_{\parallel}^2}}\approx \gamma_{\parallel} 
\quad \textrm{and} \quad \sin \alpha'= \frac{\gamma_{\perp}}{\gamma} \approx \frac{\gamma_{\perp}}{\gamma_{\parallel}}.
\end{equation}  As the transverse
component of the momentum decreases, due to the decay of the 
magnetic field, the pitch angle decreases, which makes the electron's 
momentum more aligned with the local magnetic field.

Knowing the electrons' energy distribution, the emission at any given 
frequency and time can be calculated using the synchrotron 
spectrum:

\begin{equation} 
F_{\nu} = F_{\nu_i} \left\{ \begin{array}{ll}
(\nu/\nu_a)^2 (\nu_a/\nu_i)^{1/3} & \textrm{$\nu < \nu_a$} \\
(\nu/\nu_i)^{1/3} & \textrm{$\nu_a < \nu < \nu_i$} \\
(\nu/\nu_i)^{-(p-1)/2} & \textrm{$\nu_i < \nu < \nu_c$}, \\
\end{array} \right.
\end{equation} for the case $\nu_a < \nu_i < \nu_c$, where $\nu_a$ 
is the self absorption frequency and it is obtained using 
equation (52) of Panaitescu \& Kumar (2000) (see, e.g., Katz \& Piran
1997, Sari \& Piran 1999b).  
The characteristic synchrotron frequencies 
are obtained from the corresponding electrons LFs:

\begin{equation}
\nu_{i,c}= \frac{eB\gamma_{i,c}^2 \Gamma}{2 \pi m_e (1+z)}\sin \alpha',
\end{equation} where $\gamma_{i,c}$ and $\sin \alpha'$ are given by 
(2).  The observed peak flux is

\begin{equation}
F_{\nu_i}= \frac{(1+z)\sqrt{3}e^3 N_e B \Gamma}{4 \pi d_L^2(z) m_e}\sin \alpha',
\end{equation} where $d_L$ is the luminosity distance, $N_e$ 
is the number of radiating electrons (which in this scenario is 
constant), $z$ is the redshift and 
$e$ is the electron's charge.  

For the SSC case, the flux peaks at $\nu_i \gamma_i^2$
with magnitude $\tau_e F_{\nu_i}$, where $\tau_e=N_e \sigma_T/(4 \pi R^2)$ is the 
optical depth to electron scattering.  We will calculate SSC 
emission for photons above $\nu_a$.  We will use the same 
synchrotron piece-wise spectrum, which is just a very crude 
approximation.  

\subsection{Temporal and spectral properties}

For synchrotron emission of a cooling ember undergoing AE, 
we find:

\begin{equation}
F_{\nu_i} \propto t^{-3}(t^{-3/2}), \quad \nu_{i,c} \propto t^{-3}(t^{-3/2}), 
\quad \nu_a \propto t^{-12/5}(t^{-9/5})
\end{equation} and 

\begin{equation}  
F_{\nu} \propto \left\{ \begin{array}{ll}
t^2 (t^{2}) \nu^2 & \textrm{$\nu < \nu_a$}\\
t^{-2}(t^{-1}) \nu^{1/3} & \textrm{$\nu_a < \nu < \nu_i$} \\
t^{-3(p+1)/2}(t^{-3(p+1)/4}) \nu^{-(p-1)/2} & \textrm{$\nu_i < \nu <\nu_c$,} \\
\end{array} \right.
\end{equation} where the time dependences are reported for the thin ejecta shell
and parenthesis are used for the thick ejecta shell.

Using the same notation as above, for SSC emission, we find:

\begin{equation}
F_{\nu_i \gamma_i^2} \propto t^{-5}(t^{-7/2}), \quad \nu_{i,c}\gamma_{i,c}^2 \propto t^{-3}(t^{-3/2}) 
\end{equation} and 

\begin{equation} 
F_{\nu} \propto \left\{ \begin{array}{ll}
t^{-4} (t^{-3}) \nu^{1/3} & \textrm{$\nu \lae \nu_i \gamma_i^2$} \\
t^{-(3p+7)/2}(t^{-(3p+11)/4}) \nu^{-(p-1)/2} & \textrm{$\nu_i \gamma_i^2 \lae \nu \lae \nu_c \gamma_c^2$.} \\
\end{array} \right.
\end{equation} 

One can see that for the synchrotron case, the flux decays 
rapidly for $\nu_i < \nu < \nu_c$; for the SSC case, 
the flux decays even more rapidly for
$\nu_i \gamma_i^2 \lae \nu \lae \nu_c \gamma_c^2$.  For both
cases, the peak frequencies of the spectrum also show a 
fast decrease with time.  Also, for both cases, 
the thin ejecta case gives a faster time decay, since the 
shell spreads significantly, allowing the ejecta to 
cool faster.     

To compare our theory with the observations, we provide 
relations between the temporal 
decay index $\alpha$ and the spectral index $\beta$ in 
Table 1, using the convention $F_{\nu} \propto t^{-\alpha}\nu^{-\beta}$.  

\begin{table}
\begin{center}
\begin{tabular}{ccc}
\hline
& Synchrotron & SSC \\
& $\nu_i < \nu < \nu_c$ & $\nu_i\gamma_i^2 \lae \nu \lae 
\nu_c \gamma_c^2$ \\
\hline \hline
Thick ejecta & \raisebox{-2.0ex}{$\alpha=1.5\beta+1.5$} & \raisebox{-2.0ex}{$\alpha=1.5\beta+3.5$}\\
($\Delta'=\Delta_0$) & & \\
\hline
Thin ejecta & \raisebox{-2.0ex}{$\alpha=3\beta+3$} & \raisebox{-2.0ex}{$\alpha=3\beta+5$} \\
($\Delta'=R/\Gamma$) & & \\
\hline

\end{tabular}
\end{center}
\caption{Closure relations between $\alpha$ (decay index) and 
$\beta$ (spectral index) for a cooling ember undergoing AE ($t_0=t_c$).}
\end{table}

To summarize, the emission from an adiabatically cooling source 
has the following properties: 

\begin{enumerate}
\item Its spectral index must be equal to the one at the end 
of the prompt emission phase of the gamma-ray burst,  
$\beta_{\gamma}$.  

\item The temporal decay index must obey one of the 
closure relations in Table 1.

\item The peak frequency of the spectrum should decrease
with time as predicted in (6) and (8). 

\item After some time, on the order of $t_c$ (defined in \S3.5),
the spectrum should have an exponential cut-off at frequencies greater than 
the cooling frequency (\S3.2). 

\end{enumerate}

Points (ii) and (iii) have to correspond to the same 
radiation mechanism (synchrotron or SSC) and the same
ejecta width case (thin or thick).

If one were to consider the electrons' energy as given by the
adiabatic expansion of an ideal relativistic gas, instead of using the 
methods of adiabatic invariance, then $\gamma \propto V^{-1/3}$ (\S2,
see also \S3 of M\'esz\'aros \& Rees 1999).
In this case, the velocity distribution of the electrons during the 
adiabatic expansion phase is isotropic, therefore, $\sin \alpha'$ is 
a time independent constant of order unity.  We calculate the temporal 
decay indices as done above and find the following results.  For  
synchrotron: $\alpha = 2.3 \beta +1$ and $\alpha = 4 \beta + 2$, 
and for SSC: $\alpha = 3.7 \beta +3$ and $\alpha = 6 \beta + 4$, 
for the thick and thin ejecta cases, respectively.  The difference 
in the temporal decay indices for the synchrotron case compared 
to the ones on Table 1 is $\lae 20\%$ for $\beta \in [0.5-2]$.  
Because SSC has a stronger dependence on $\gamma$, the 
difference we find in $\alpha$ is larger.

So far, we have considered only the flux-freezing condition to 
prescribe the evolution of the magnetic field, but we can also 
determine the magnetic field using the equipartition 
consideration, i.e. the energy density in the magnetic field is a 
constant fraction of the electrons' internal energy density
(Sari, Piran \& Narayan 1998).  
To obtain this last quantity, one needs to know $\gamma$, 
which could be obtained either using the adiabatic invariance 
methods or by the ideal gas law - as mentioned 
in the last paragraph.  But for the equipartition consideration
we will only consider the ideal gas law, because if there is 
a mechanism that maintains an equipartition between magnetic 
energy and electron energy, then that same process is also 
likely to keep different components of electron momentum 
coupled and that will lead to an ideal gas expansion law 
for electrons.  Therefore, the magnetic field in this 
case is given by $B^2 \propto V^{-4/3}$, where $V \propto R^2 \Delta'$.  
The synchrotron and SSC emission decays, for the thick case, are both steeper 
by $0.3 \beta + 0.3$ than the ones presented on the last paragraph, 
but both thin cases remain unchanged. 

\subsection{Large Angle Emission}

If the central engine switches off abruptly and the 
gamma-ray producing ejecta has a opening angle $\theta_j$, 
such as $\theta_j > \Gamma^{-1}$, Large Angle Emission (LAE)
will be also present (Fenimore \& Sumner 1997, Kumar \& Panaitescu 2000).
The LAE flux declines as $\alpha=2+\beta$ and the peak frequency of the spectrum
decays as $t^{-1}$.  Therefore, the AE flux 
generally\footnote{Except when $\beta < 1$ for the AE case 
of synchrotron emission from a thick ejecta shell.} decays 
faster than LAE's and the AE's peak frequencies always decrease 
faster than LAE's.   

The time-scales for these two phenomena, LAE and AE, are essentially 
the same, they are set by 

\begin{equation}
t_c=\frac{R}{2\Gamma^2}. 
\end{equation}  LAE and AE start at the same time, $t_0$, 
and same site, $R_0$: right after the central 
engine has switched off, and the fluxes decline with time as:

\begin{equation}
F_{\nu}=F_0 \Big(1 + \frac{t - t_0}{t_c}\Big)^{-\alpha},
\end{equation} where $F_0$ is the flux at $t_0$ and $\alpha$
is the decay index of either LAE or AE.  The shape of the 
LC depends on the values for $t_0$ and $t_c$ (Figure 1);
for \S3.4, $t_0 = t_c$.  The case $t_0 < t_c$ is unphysical, 
since it implies that when AE starts, the shell's electrons 
have not yet cooled substantially, i.e. the shell's radius 
hasn't doubled.

\begin{figure}
\centerline{\hbox{\includegraphics[width=10cm,angle=0]{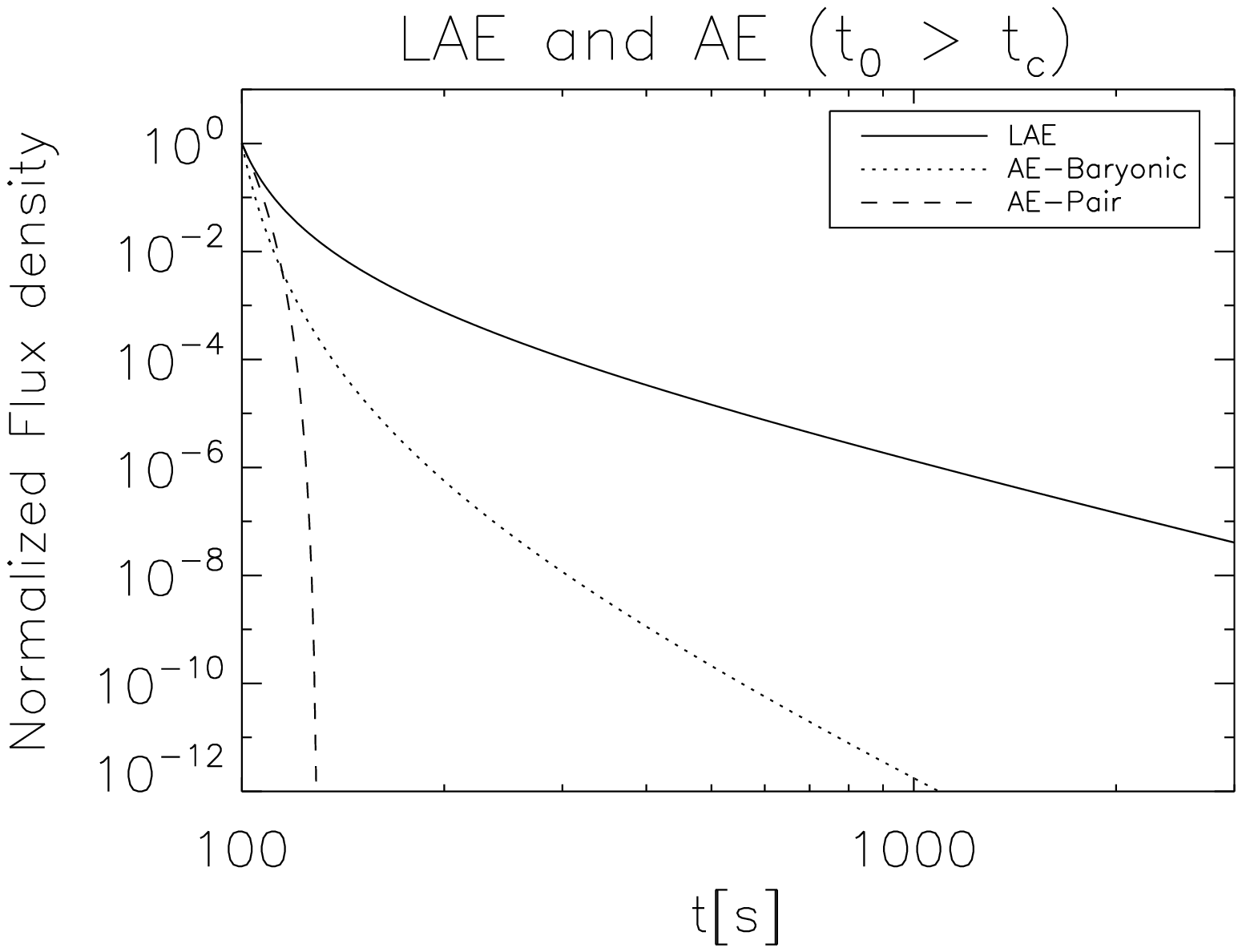}}}
\centerline{\hbox{\includegraphics[width=10cm,angle=0]{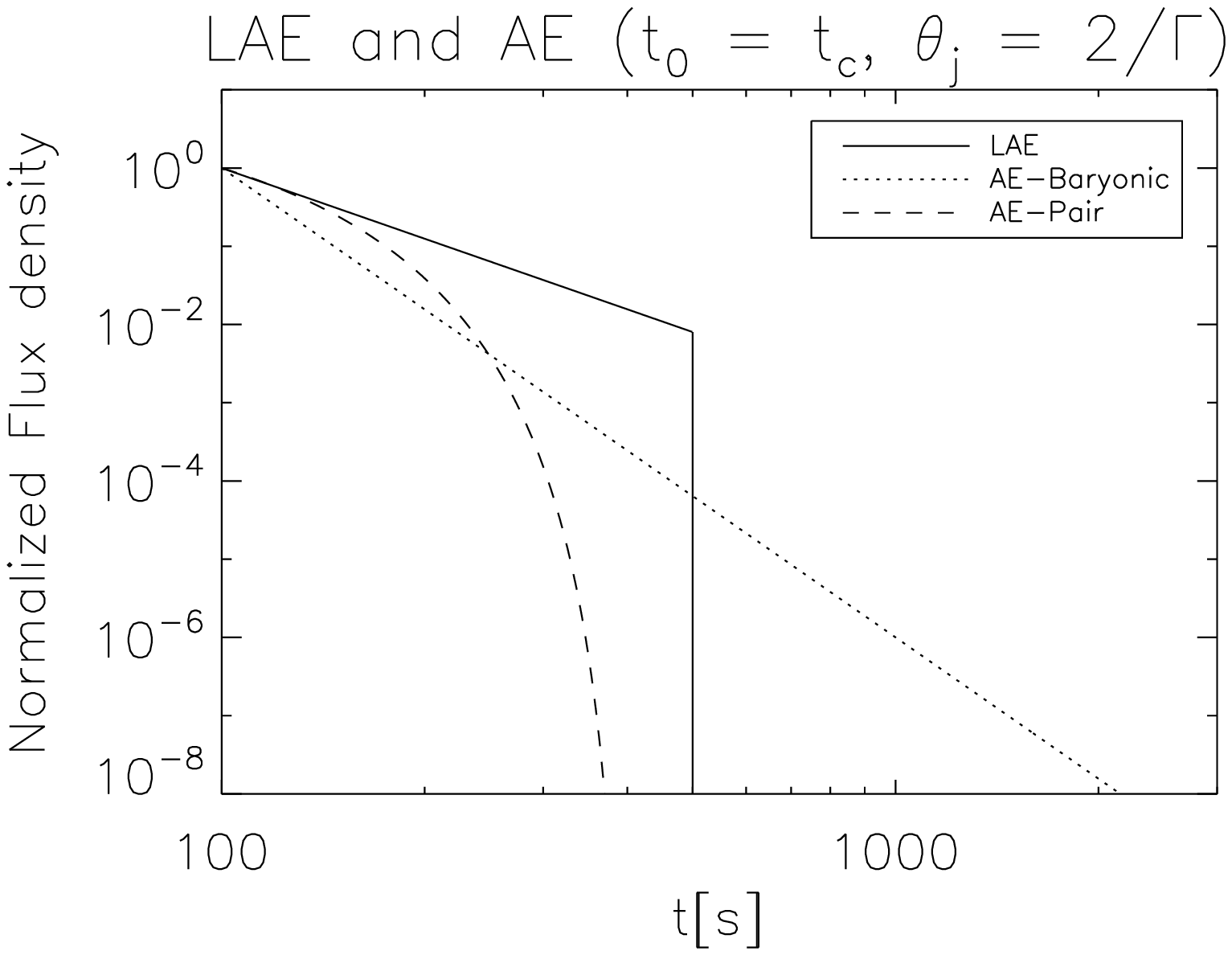}}}
\caption {The normalized flux density, equations (11) and (12), 
plotted vs. observed time, assuming that the observed frequency $\nu$ is 
always between $\nu_i$ and $\nu_c$ (if $\nu_c < \nu$, then there is no
AE, only LAE if $\theta_j > \Gamma^{-1}$, see \S3.2). This emission is produced by the 
last ejected shell, because contributions from previously ejected shells would 
be buried in the emission of subsequent shells, since both LAE and AE 
decay very fast. {\bf Top.} Using $t_0=100s$ and $t_c=10s$.  
The LAE and AE-Baryonic 
decay indices correspond to $\alpha=(3,6)$, 
respectively, and the AE-Pair has $\delta=8$ ($\beta=1$, using AE:
Synchrotron - thin ejecta). {\bf Bottom.} 
Using $t_0=t_c=100s$ and the same $\alpha$'s and $\delta$ as above.  
In this illustrative example we have $\theta_j=2/\Gamma$, 
therefore, LAE dominates over AE until $t_j=500s$, when AE-Baryonic 
takes over. This break in the LC from the LAE to the AE 
power law decay (which should be a smooth transition and it is done
in the figure for display purposes) has never been observed.}
\end{figure}

If $\theta_j \lae \Gamma^{-1}$, then 
there will be no LAE, so AE would be the only emission present 
after the central engine turns off. 
On the other hand, if $\theta_j > \Gamma^{-1}$, then  
LAE will dominate over AE (see footnote 2).  
LAE will cease with the detection of the last photons 
coming from $\theta_j$ and, at this time, the flux will smoothly 
become dominated by the AE emission, i.e. there will be a break in the LC 
to the power law decay for AE (Figure 1: Bottom).  
The photons from $\theta_j$ will arrive at a time 
$t_j \approx t_0 + R\theta_j^2/2 = t_0 + \theta_j^2\Gamma^2 t_c$.

\subsection{Electron-positron pair-enriched ejecta}

When the ejecta cools by adiabatic expansion, the thermal 
energy of the protons and electrons is converted back to 
bulk kinetic energy of the shell, increasing $\Gamma$.  Even 
in an extreme case where all the electrons' energy goes 
into the shell expansion, $\Gamma$ increases 
only by a factor of $\sim2$, if the protons and electrons energy 
is $\lae m_p c^2$.\footnote{We have assumed a co-moving observer
sitting in the middle of an infinite parallel shell 
that sees the left and right halves of the shell move away from him. 
Assuming the electrons' LF in the shell rest frame is $10^3$ (and that 
the protons are essentially cold since the heating mechanism energized 
all particles equally), the LF 
of the shells would be $1 + 10^3 m_e/m_p = 1.5$.  An observer far 
away would mainly detect radiation from the half moving towards him, 
since the radiation is beamed.  This observer would see that the LF 
of this half has increased by a factor of $\sim2$.}  Therefore, the effect of this 
change to the observed flux is less than a factor of $2$, 
a relatively small effect.  For this reason, we have used a constant $\Gamma$ for 
the calculations done so far.  

On the other hand, if the ejecta consists of $e^{\pm}$ pairs, 
then the increase in $\Gamma$ during the adiabatic expansion
would be considerable, 
and it would scale as $\propto \gamma^{-1}$ (the observed energy in 
the shell is a constant and scales as $\propto \gamma \Gamma$).  
Since $\Gamma$ increases, then the observed time is 
$t-t_0= \int^R_{R_0}\,dR/(2 \Gamma^2)$.   
For the thin ejecta case, we find:

\begin{equation}
F_{\nu}=F_0 \Big(1-\frac{t-t_0}{3 t_c}\Big)^{\delta}, 
\end{equation} with characteristic frequencies $\propto [1-(t-t_0)/(3 t_c)]^{4}$, 
where $\delta =(4 \beta + 4,4 \beta + 10)$ for the synchrotron
($\nu_i < \nu < \nu_c$) and SSC ($\nu_i \gamma_i^2 \lae \nu \lae \nu_c \gamma_c^2$)
cases, respectively.  The time decay index, for the $t_0 = t_c$ case, is 
$\delta \frac{(t/t_0)}{(4 - t/t_0)}$, therefore, the LC steepens continuously.  It 
decays even faster than the AE-Baryonic case, since the observed time gets compressed 
because $\Gamma$ is increasing considerably.  
If $\theta_j > \Gamma^{-1}$, then LAE prevents it from steepening more 
than $2 + \beta$: completely taking over the emission since essentially $t_0$
(Figure 1: Bottom).          

\subsection{Reverse Shock emission}

In this short subsection, we explore the possibility that the 
GRB ejecta, that just produced the prompt emission, interacts 
with the interstellar medium (ISM) and a reverse shock (RS)
crosses it.  The ejecta cools adiabatically after the RS 
has passed through it and we assume that it follows the 
Blandford-McKee self similar solution (Blandford \& McKee 1976), 
during which $\Gamma$ decays in time.  

Using the same methods as in the previous subsections, we 
can calculate the LF of the electrons in the ejecta after 
the passage of the RS (see Appendix B).  We find that the RS flux decays 
as $\propto t^{-411(p+1)/568}=t^{-2.53}$ 
(synchrotron emission: $\nu_i < \nu < \nu_c$, for a thin shell,
using $B_{\perp}$ and $p=2.5$), which gives the closure 
relation: $\alpha = 1.45 \beta + 1.45$ (and $\alpha = 1.45\beta +1.67$ 
for SSC).  If we determine the electrons' LF 
using the ideal gas law,  
then the RS synchrotron flux would decay as $\propto t^{-(20p+7)/24}=t^{-2.38}$ 
(for the same case as above), 
which is still steeper than the $\propto t^{-(73p+21)/96}=t^{-2.12}$ derived by 
Sari \& Piran (1999a), where they used $\gamma \propto V^{-1/3}$ and 
the equipartition consideration.

\section{Application to the GRB early ``afterglow''}

AE, together with LAE, dictates the emission of the source 
after the central engine has completely turned off.  In this 
section, we will determine if the early x-ray steep decay
observed by {\it Swift} obeys our theoretical LCs for AE and LAE
\footnote{If the prompt emission is attributed to synchrotron, 
then, for some fraction of the parameter space, the radiative cooling 
timescale, $t_{rad}$ is less than $t_c$.  
However, shortly after the onset of the adiabatic expansion, 
$t_{rad} > t_c$, since the magnetic field decays rapidly with 
the expansion of the ejecta.  For the SSC case, $t_{rad} \gae t_c$
is very likely at the onset of the adiabatic 
expansion, making the radiative cooling unimportant.}.
We will do this for each one of the cases considered in the previous 
section.  

The early x-ray data shows a single power law decay
with $3 \lae \alpha \lae 5$ (Nousek et al. 2006, O'Brien et al. 2006, 
Willingale et al. 2007).  With this information, the $t_0 > t_c$ case 
can be ruled out, since, for this case, 
the theoretical shape of the LCs for LAE and AE-Baryonic is 
inconsistent with the early x-ray observations and the AE-Pair 
LC decays extremely fast (Figure 1: Top). Therefore, 
we focus on the $t_0=t_c$ case only. 

The next possibility we explore is to see if the early x-ray data 
obeys LAE or AE (from a baryonic ejecta).  
To check the validity of these two scenarios, respectively, we will take a 
sample of bursts and see how many 
cases are possibly consistent with either LAE or AE.   

Our sample consists of 107 GRBs for which their spectral index 
during the early x-ray decay ($\beta_x$) and their temporal 
decay index during this phase have been previously determined
(the sample of Willingale et al. 2007). 
We first select the bursts for which $\beta_{\gamma}=\beta_x$, which 
cuts down the sample to 55 bursts.  Eight of these 
bursts show strong spectral evolution, 
inconsistent with LAE and with AE  (Zhang et al. 2007: 
Zhang et al's sample essentially contains all our sample), 
which leaves us with 47 bursts.
Moreover, we check how many of these satisfy
the relations between $\alpha$ and $\beta$ for LAE or AE 
(Table 1) within about a 90\% confidence level, and that 
narrows down the sample to 20 bursts.
In conclusion, only a small percentage of the sample, 19\%,
is consistent with LAE or AE,  which 
leads us to suggest that, for most bursts, the early x-ray 
data results from some other process, and the most natural 
conclusion is continued central engine activity.

It has also been claimed that the $\gamma$-ray emission extrapolated
to x-ray energies, together with the early x-ray data, can be well fitted 
with a falling exponential followed by a power law (O'Brien et al. 2006).
At first, this could be thought to be explained by a pair ejecta with 
$\theta_j \lae \Gamma^{-1}$ undergoing AE, since its LC also steepens continuously 
(Figure 1: Bottom).  However, for 3 bursts that show this continuous steepening: GRB 050315 
(Vaughan et al. 2006, Lazzati \& Begelman 2006), GRB 050724 (Barthelmy et al. 2005) 
and GRB 060614 (Mangano et al. 2007), 
the theoretical LC decays too fast and can't fit the observed early x-ray 
LC.  Therefore, we rule out the possibility
that the observed early x-ray decay is from a pair ejecta undergoing AE.

Finally, we use our sample to test if the observed early x-ray steep decay 
is consistent with the closure relations derived for the RS (\S3.7).  In 
this case, the condition $\beta_{\gamma} = \beta_x$ is not necessary, 
therefore we start with the entire sample: 107 bursts.  We eliminate
19 of these, which show strong spectral evolution, inconsistent with RS.  
Out of the remainder, only 26 bursts (24\%) are possibly consistent with 
the RS (16 of these are simultaneously consistent with LAE and AE). However, 
this scenario could be ruled out since it 
would be difficult to explain the connection observed in the LC 
between the prompt emission and the early x-ray data.  The only way 
they could be smoothly connected is if the prompt gamma-ray emission 
is produced also by the RS, for which we lack evidence.

\section{Discussion: The central engine}

The results of the last section lead us to believe that the observed 
early x-ray decay for $\gae$ 70\% of GRBs is produced by the rapidly 
declining continued activity of the central engine,
if we assume that there is a one-to-one correspondence
between the temporal behavior of the central engine activity  
and the observed emission (Figure 2).

There are two other arguments that support the idea that, for 
most bursts, neither LAE nor AE might be consistent with the observed
early x-ray decay.  First, for some bursts, a break frequency 
has been seen passing through the x-ray band during the early 
steep decay, and it evolves faster than the $\propto t^{-1}$ 
expected in LAE: $\propto t^{-2}$ for GRB 060614 (Mangano et 
al. 2007\footnote{Mangano et al. mention in 
their work that this could be attributed to the $\nu_c$
decrease due to adiabatic cooling of shock heated shells after 
an internal shock.}) 
and $\propto t^{-4} \sim t^{-3}$ for GRB 060904A (Yonetoku et al. 2008).
Second, if AE is entirely responsible for this 
phase, then $\theta_j$ has to be very small.  If this is the case, 
then we should have observed the edge of the jet (a jet break 
with a $\propto t^{-p}$ optical LC) very early on.  
After inspecting many 
optical LCs (Butler \& Kocevski 2007, Liang, E.-W. et al. 2007, Liang, 
E.-W. et al. 2008, Melandri et al. 2008, Panaitescu \& Vestrand 2008), 
we can conclude that most bursts with available early optical 
data don't show this expected jet break starting at 
very early times, i.e. $t <$ a few hours. 

One way that we might have missed these early jet breaks could be
explained by the ``porcupine'' model.  In this model, the central 
engine ejects many very small angle ($\theta_j \lae \Gamma^{-1}$) jets. One of 
these jets, directed towards the observer, produces the gamma-ray emission. 
The central engine shuts off, and then the AE emission is produced.  After a short time, 
all these small jets combine and give rise to a single jet with 
$\theta_j > \Gamma^{-1}$ that interacts with the interstellar medium,
giving rise to a forward shock (optical afterglow). This model leaves 
no sign of a jet break.  Another possible explanation for 
the lack of a very early jet break is 
that there is energy injection to the forward shock, making the $\propto t^{-p}$ optical 
LC more shallow.  This last scenario is unlikely, as a large amount 
of energy is required - more than a factor of $10$ increase - to make 
a $\propto t^{-p}$ LC as shallow as $\sim t^{-1}$, which is the usual 
observed optical LC decay.

It has been suggested that the observed early x-ray decay is produced by the 
forward shock (FS) driven by the ejecta interacting with the ISM (Panaitescu 2007).  
This scenario has problems explaining the smooth temporal connection in the 
LC between the prompt emission and the early x-ray steep decay. 

The model presented in this paper has several uncertainties.  For example, 
we are not specifying how the magnetic field is generated during the 
prompt emission.  We are assuming: (i) that the magnetic field 
coherence length-scale is larger than the electron gyro-radius, 
based on the observations (see Appendix A), and (ii) that there 
are no collective plasma effects that randomize the 
electrons' velocity.  These assumptions allow us to use the 
adiabatic invariant presented in \S2.  
If, for some reason these conditions are violated, then 
the adiabatic invariance of electron magnetic moment can't be used.  
However, the light-curve from an expanding shell is similar whether
we follow electron cooling via adiabatic invariance or ideal gas
law and therefore, the main conclusions we have presented
here are unchanged.  Another assumption made in the 
application of this model is that {\it Swift}'s x-ray telescope band lies between
$\nu_i$ and $\nu_c$ (in that order), which can be inferred from the 
spectral information of most bursts during this phase.  If, 
however, $\nu_c < \nu_i$, then there will be no emission coming 
from the part of the shell that lies within an angle of $\Gamma^{-1}$
to the observer line of sight - the only emission would be from LAE 
(if $\theta_j > \Gamma^{-1}$).

\begin{figure}
\centerline{\hbox{\includegraphics[width=6cm,angle=0]{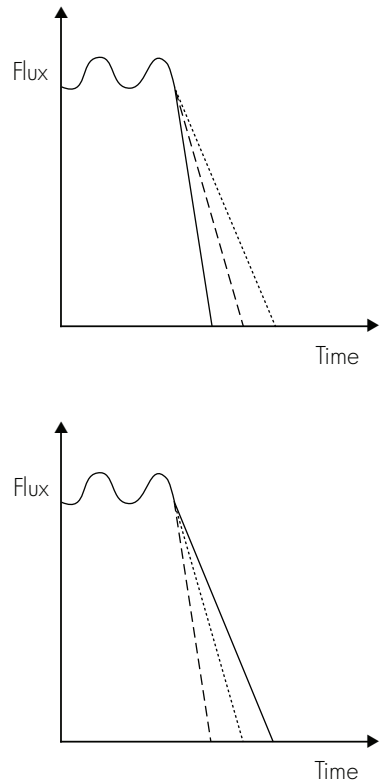}}}
\caption {Two possible scenarios for the contribution of LAE (dotted),
AE (dashed) and the rapidly declining central engine activity (solid), 
as seen in the x-ray band.  The two ``humps'' represent the $\gamma$-ray
detection (from {\it Swift} BAT), extrapolated to the x-ray band, attributed
to activity of the central engine. {\bf Top.} The case for which the central 
engine activity drops extremely fast and LAE and AE appear. Only $\sim 20\%$ of our sample 
is possibly consistent with this scenario. {\bf Bottom.} Our preferred scenario, 
where the central engine activity is the dominant contribution and 
it decays slower than the theoretical LCs for LAE and AE.}
\end{figure}

\section{Conclusions}

We have explored the situation in which the central engine shuts
off and the ejecta cools via adiabatic expansion (AE).  We have
derived and discussed this emission's temporal and spectral
properties using a new treatment for the micro-physics of the
AE: the adiabatic invariant $\gamma_{\perp}^2/B$, describing the
electron momentum normal to the magnetic field (see summary
in \S3.4).  At the onset of the adiabatic expansion, as $B$ 
decays rapidly, this component of the  momentum decreases 
while the parallel one remains unchanged, making the electrons 
more and more aligned with the local magnetic 
field as the ejecta expands.  The adiabatic invariant 
enables us to calculate the electrons' energy for a collisionless
magnetized plasma, if no other collective plasma effects that randomize
the electrons' velocity are present. 

In regards to the central engine activity, we can draw a 
conclusion:  The fastest way that the observed flux 
can decline after the central engine shuts off 
is set by the Large Angle Emission (LAE) and the Adiabatic 
Expansion cooling (depending on the value of $\theta_j$).

The early x-ray steep decay shown in most of {\it Swift} 
bursts has been attributed to LAE.  In this paper, we 
consider both AE and LAE for the very early x-ray data.  LAE 
and AE both start with the assumption that the central 
engine shuts off abruptly.  Only $\sim 20\%$ of our sample 
of 107 bursts is possibly consistent with either LAE or AE, 
thereby suggesting that the observed early 
x-ray steep decay for a large fraction of bursts might be 
produced by the rapidly declining
continuation of the central engine activity.  

The component of the electron's momentum parallel to the magnetic 
field is unconstrained by the adiabatic invariance. This component 
would probably cool via Inverse Compton scattering 
with synchrotron photons.  Another possibility is
that the electrons are scattered by small scale 
fluctuations in the magnetic field, which would effectively
couple the parallel and perpendicular components of their 
momentum, resulting in an adiabatic cooling similar to that 
in the ideal gas case, \S2 (personal communication, Granot).

Any process that does not rely on the central engine 
activity to explain the observed early x-ray steep decay 
(i.e. RS, FS) has problems explaining the smooth 
temporal connection observed in the LC between the prompt emission and 
the early x-ray steep decay. 

\section*{Acknowledgments}
This work is supported in part by grants from NSF (AST-0406878) and NASA
Swift-GI-program. RBD thanks Jessa Hollett for her support during the 
writing of this manuscript and for her help on Figure 2.  RBD also thanks
Rongfeng Shen, Ehud Nakar, Jonathan Granot and Milo\v{s} Milosavljevi\'c for helpful comments 
about this project.  We thank the referee for helpful comments and 
constructive suggestions.  



\renewcommand{\theequation}{A-\arabic{equation}}
\setcounter{equation}{0}  
\section*{APPENDIX A: Electron gyro-radius vs. Magnetic field length-scale}  

The observed peak energy in the prompt phase of a GRB 
is given by (4) and it is 
$\nu_i = (1.15 \times 10^{-8}eV)B \gamma_i^2 \Gamma (1+z)^{-1}$,
assuming synchrotron emission.
The value of $\Gamma$ can be constrained to be 
a few hundred.  There is a wide 
range of allowed values for $\gamma_i$.  For 
$\gamma_i=10^3$, $\nu_i=100keV$, 
$\Gamma = 100$ and $z=1$, then $B = 2 \times 10^5 G$, and in 
that case, the electrons' gyro-radius is $r=m_e c^2 \gamma_i/(eB) \approx 10 cm$.

If the magnetic field responsible for the prompt emission 
is the frozen-in field from the central explosion, then 
it decays on a length-scale on the order of the source size, which is
much larger than the electrons' gyro-radius and the adiabatic invariant presented 
in \S1 can be used.
  
If the field is produced locally (e.g. by the Weibel instability), then 
we need to estimate its coherence length and compare it with
the electrons' gyro-radius.  For instance, magnetic field generated 
by the Weibel instability will have a coherence length on the order of the 
plasma length $\lambda_B=c/\omega_p$, where 
$\omega_p=(4\pi e^2 n/m_e)^{1/2}=6\times10^4 (n)^{1/2} s^{-1}$ 
and $n$ is the co-moving electron number density in units of $cm^{-3}$.
For the prompt emission, recent studies have shown that the radius 
of emission is on the order of $10^{15-16} cm$ or 
even larger (Kumar et al. 2007, Racusin et al. 2008, Zou, Piran \& Sari 2009, 
Kumar \& Narayan 2008), 
therefore $n=5 \times 10^{5-8} cm^{-3}$
(see footnote 1), which gives $\lambda_B=20-700 cm$, 
making $\lambda_B$ larger than $r$ 
by at least a factor of $2$.  Moreover, this magnetic field 
decays extremely fast in time (in about 
$\omega_p^{-1}$), which would be less than $10^{-8} s$ in the 
source co-moving frame or $10^{-10} s$ in the observer frame.
This locally generated magnetic field cannot be responsible 
for the prompt emission, unless it is sustained for at least
$\sim 1s$ (the co-moving time-scale of a few mili-second prompt pulse), 
which would require a much larger $\lambda_B$ (Keshet et al. 2008 mention
that the field must persist over $10^{10} \lambda_B$ downstream, which in
this case would be $\sim 10^{12} cm$).  
Therefore, it is safe to assume that even if the field is generated 
locally $r << \lambda_B$, allowing us to use the adiabatic invariant.

The prompt emission phase could also be attributed to the 
SSC emission, which requires smaller values for $\gamma_i$ and 
$B$.  For $\gamma_i=10^2$, then $B=2 \times 10^3 G$, 
which gives $r \approx 100 cm \sim \lambda_B$.  But as mentioned above, 
the field has to be coherent on length-scale $\sim 10^{12} cm$ in order 
that it does not decay away on time $\lae 1s$.

\renewcommand{\theequation}{B-\arabic{equation}}
\setcounter{equation}{0}  
\section*{APPENDIX B: Reverse Shock emission calculation}

After the RS has crossed the ejecta, it follows roughly the
Blandford-McKee self-similar solution (Blandford \& McKee 1976), 
in which the bulk LF and pressure of the shocked ISM are 
given by:

\begin{equation}
\gamma(t,r) = \gamma(t)\chi^{-\frac{1}{2}}, 
\quad P(r,t)=4 m_p c^2 n [\gamma(t)]^2\chi^{-\frac{17-4s}{12-3s}},
\end{equation} where $\gamma(t) \propto R^{-(3-s)/2}$ is the LF of 
material just behind the shock, $n \propto R^{-s}$ is the ISM particle number 
density, $\chi$ is the similarity variable and $m_p$ is the 
proton mass (see, e.g, Sari 1997).

Let us assume that the ejecta is at $\chi_{ej}$ and it has a 
pressure $P_{ej}$ and a LF $\Gamma_{ej}$, 
which - because of pressure and velocity equilibrium at the 
contact discontinuity - should be the same as the bulk 
LF and pressure of the shocked ISM at $\chi_{ej}$ (B-1).  
The pressure in the 
ejecta for the thin case ($\Delta' = R/\Gamma_{ej}$) is 
given by $P_{ej} \propto V^{-1}\gamma_{\perp}$, 
where $V \propto R^2 \Delta'$, and using the 
adiabatic invariance (1), we can use 
$\gamma_{\perp} \propto B_{\perp}^{1/2}$, where $B_{\perp}$
is given in \S3.1.  Therefore, we have 
$P_{ej} \propto R^{-4} \Gamma_{ej}^{3/2}$ and using (B-1) and 
the contact discontinuity equilibrium conditions we obtain:

\begin{equation}
\Gamma_{ej} = \gamma(t)\chi_{ej}^{-\frac{1}{2}}, 
\quad P_{ej}=4 m_p c^2 n [\gamma(t)]^2\chi_{ej}^{-\frac{17-4s}{12-3s}}.
\end{equation} We can solve for $\Gamma_{ej}$ 
in terms of $R$, which gives $\Gamma_{ej} \propto R^{-\frac{2(63-32s+4s^2)}{32-7s}}$.
For uniform ISM ($s=0$), we can determine the observed time by
$t-t_0= \int^R_{R_0}\,dR/(2 \Gamma_{ej}^2)$ and following 
the procedure on \S3.3, we can find that the RS shock flux decays as 
$\propto t^{-411(p+1)/568}=t^{-2.53}$ for $p=2.5$ ($\nu_i<\nu<\nu_c$).

We can also determine the electrons' LF by using the
ideal gas law.  For this case, $\gamma \propto V^{-1/3}$, 
therefore the pressure in the ejecta is $P_{ej} \propto V^{-4/3}$.
For the thin ejecta case, we can repeat the calculation 
done before (B-2), just modifying $P_{ej}$, and we obtain 
$\Gamma_{ej} \propto R^{-(7-2s)/2}$.  To calculate the 
RS synchrotron emission for uniform ISM, we can use 
$B_{\perp} \propto (R \Delta')^{-1}$ and we can 
follow the procedure on \S3.3, but as mentioned before, 
for this case $\sin \alpha' \sim 1$. We find that
the RS decays in this case as $\propto t^{-(20p+7)/24}=t^{-2.38}$
for $p=2.5$ ($\nu_i<\nu<\nu_c$).  If we determine 
the magnetic field using the equipartition 
consideration and determine the electrons' LF 
with the ideal gas law, then we get the 
same result reported by Sari \& Piran (1999a).


\begin{thebibliography}{99}
\bibitem{} Barthelmy, S.D. et al., 2005, Nature, 438, 994
\bibitem{} Blandford R. D., McKee C. F., 1976, Phys. Fluids, 19, 1130  
\bibitem{} Butler N., Kocevski D., 2007, ApJ, 663, 407
\bibitem{} Fenimore E., Sumner M., 1997, ``All-Sky X-ray Observations in the Next Decade'', 
eds. M. Matsuoka and N. Kawai, Riken Institute, Japan, p. 167 (astro-ph/9705052)
\bibitem{} Jackson J. D., 1998, Classical Electrodynamics. Wiley, New York
\bibitem{} Katz J.I., Piran T., 1997, ApJ, 490, 772  
\bibitem{} Keshet U., Katz B., Spitkovsky A., Waxman E., 2008, preprint (arXiv:0802.3217 [astro-ph])
\bibitem{} Kumar P., Panaitescu A., 2000, ApJ, 541, L51
\bibitem{} Kumar P. et al. 2007, MNRAS, 376, L57
\bibitem{} Kumar P., Narayan R., 2008, accepted to MNRAS (arXiv:0812.0021 [astro-ph])
\bibitem{} Lazzati D., Begelman M.C., 2006, ApJ, 641, 972
\bibitem{} Liang E.-W., Zhang B.-B., Zhang B., 2007, ApJ, 670, 565
\bibitem{} Liang E.-W., Racusin J.L., Zhang B., Zhang B.-B., Burrows D.N., 2008, ApJ, 675, 528
\bibitem{} Mangano V. et al., 2007, A\&A, 470, 105
\bibitem{} Melandri A. et al, 2008, ApJ, 686, 1209
\bibitem{} M\'esz\'aros P., Rees M.J., 1997, ApJ, 476, 232
\bibitem{} M\'esz\'aros P., Rees M.J., 1999, MNRAS, 306, L39
\bibitem{} Nousek J.A. et al., 2006, ApJ, 642, 389
\bibitem{} O'Brien P. et al, 2006, ApJ, 647, 1213
\bibitem{} Panaitescu A., Kumar P., 2000, ApJ, 543, 66
\bibitem{} Panaitescu A., Kumar P., 2004, MNRAS, 353, 511
\bibitem{} Panaitescu A., 2007, MNRAS, 379, 331
\bibitem{} Panaitescu A., Vestrand W.T., 2008, MNRAS, 387, 497
\bibitem{} Piran T., 2005, Reviews of Modern Physics, 76, 1143
\bibitem{} Tagliaferri G. et al., 2005, Nature, 436, 985
\bibitem{} Racusin J.L. et al. 2008, Nature, 455, 183 
\bibitem{} Rybicki G. B., Lightman A. P., 1979, Radiative Processes
in Astrophysics. Wiley-Interscience Press, New York
\bibitem{} Sari R., 1997, ApJ, 489, L37
\bibitem{} Sari R., Piran T., Narayan R., 1998, ApJ, 497, L17
\bibitem{} Sari R., Piran T., 1999a, ApJ, 517, L109
\bibitem{} Sari R., Piran T., 1999b, ApJ, 520, 641
\bibitem{} Vaughan S. et al., 2006, ApJ, 638, 920
\bibitem{} Willingale R. et al., 2007, ApJ, 662, 1093
\bibitem{} Yonetoku D. et al., 2008, PASJ, 60, 263
\bibitem{} Zhang B., Fan Y.Z., Dyks J., Kobayashi S., M\'esz\'aros P., Burrows D.N., Nousek
J.A., Gehrels N., 2006, ApJ, 642, 354
\bibitem{} Zhang B.-B., Liang E.-W., Zhang B., 2007, ApJ, 666, 1002
\bibitem{} Zou Y.C., Piran T., Sari R., 2009, ApJ, 692, 92
\end{thebibliography}
\end{document}